\documentclass[aps,showpacs]{revtex4}
\usepackage{graphicx}
\usepackage{dcolumn}
\usepackage{bm}
\usepackage{color}
\usepackage{ulem}
\newcommand{\be}{\begin{equation}}
\newcommand{\ee}{\end{equation}}
\newcommand{\bea}{\begin{eqnarray}}
\newcommand{\eea}{\end{eqnarray}}
\newcommand{\beb}{\begin{eqnarray*}}
\newcommand{\eeb}{\end{eqnarray*}}
\begin{document}

\title{Gaplessness of the Gaffnian}

\author{Th. Jolicoeur$^1$}
\author{T. Mizusaki$^2$}
\author{Ph. Lecheminant$^3$}

\affiliation{ $^1$~Laboratoire de Physique Th\'eorique et Mod\`eles Statistiques, CNRS and
Universit\'e Paris-Sud, 91405 Orsay, France}
\affiliation{$^2$~Institute of Natural Sciences, Senshu University,  Tokyo 101-8425, Japan}
\affiliation{$^3$~Laboratoire de Physique Th\'eorique et Mod\'elisation, CNRS and
Universit\'e de Cergy-Pontoise, 95302 Cergy-Pontoise, France}

\date{June 23rd, 2014}
\begin{abstract}
We study the Gaffnian trial wavefunction proposed to describe fractional quantum Hall correlations
at Bose filling factor $\nu =2/3$ and Fermi filling $\nu =2/5$.  A family of Hamiltonians interpolating between
a hard-core interaction for which the physics is known and a projector whose ground state is the Gaffnian
is studied in detail. We give evidence for the absence of a gap by using
large-scale exact diagonalizations in the spherical geometry. This is in agreement with recent arguments based
on the fact that this wavefunction is constructed from a non-unitary conformal field theory.
By using the cylinder geometry, we discuss in detail the nature of the underlying minimal model and
we show the appearance of heterotic conformal towers in the edge energy spectrum where left and right
movers are generated by distinct primary operators.
\end{abstract}
\maketitle
\section{Introduction}
Coulomb interactions between electrons in the lowest Landau level (LLL) lead to the appearance
of the fractional quantum Hall effect (FQHE). Since the kinetic energy is quenched by the magnetic field,
the nature of phases is entirely dictated by the interactions and is not amenable to standard many-body treatments.
Hence the use of trial wavefunctions has been the key to unlock the FQHE physics. The most experimentally prominent
quantum Hall fraction at LLL filling factor $\nu =1/3$ has been explained by Laughlin by means of an explicit
first-quantized wavefunction~\cite{Laughlin83}. Some successful wavefunctions have been introduced
 to describe other fractions in the picture of ``composite fermions'' (CF)~\cite{JainBook}. Another line
of attack uses conformal field theory (CFT) to produce candidate wavefunctions by computing expectation values
of CFT vertex operators. The prime example in this family is the Moore-Read ``Pfaffian'' which is build from
an Ising-like theory~\cite{Moore91}. It is a promising candidate~\cite{RMP08,Kumar2010}
 to describe the FQHE observed in the second Landau level
at filling $\nu=5/2$. To assess the relevance of a given FQHE wavefunction, there are a handful of tools
at our disposal. If the function has an explicit analytical expression one may compute observables by
Monte-Carlo evaluation of expectation values. This is the case of the CF wavefunctions.
It is also possible to test these predictions with exact diagonalization (ED) of systems with small numbers of particles
that can be treated without ad-hoc approximations. Some of the trial wavefunctions have also a simple property~:
they are exact ground states of some local positive-definite quantum Hamiltonians. This last case is relevant
to wavefunctions that have special vanishing properties when two or more particles come close.
The important quantities that should be extracted from a given trial state  are the charge and statistics of the
quasiparticles, the edge mode characteristics with their scaling exponents. In principle they can be measured
by using various experimental techniques. For example  tunneling measurements give access to scaling exponents, 
 interferometry may lead to the braiding statistics of quasiparticles, noise measurements can give the charge
and thermal transport can also be used to find exponents.
 
In the CFT approach, it is immediately clear what are the physically relevant theories from which one can
construct a trial wavefunction. The Laughlin state can be derived from a simple compact boson theory with $U(1)$
symmetry. The Pfaffian state is obtained by adding an extra Ising CFT on top of a similar boson theory.
It can be seen as the first member of a set of so-called Read-Rezayi~\cite{RR} states that involve $Z_k$ parafermions.
The $Z_k$ parafermions belong to the series of $\mathcal{W}_k$ minimal models. Such models are indexed by two integers
$\mathcal{W}_k(p,p^\prime)$ and the parafermions correspond to the subset $\mathcal{W}_k(k+1,k+2)$.
There is evidence
that they play a role in some FQHE states. Indeed the state observed at filling $\nu=2+2/5$ in some samples
has been suggested to be related by particle-hole symmetry in the second Landau level  to the
$\nu=3/5$ fermionic $Z_3$ parafermionic state. In the realm of the bosonic FQHE which may be
relevant to ultracold atomic gases~\cite{Regnault03,Chang05}, the sequence of states with filling $\nu =k/2$ for $k=1,2,3,4,\dots$
may be the ground state after melting of the vortex lattice. These $W_k$ models do not exhaust the list~\cite{ginsparg,MalteBook}
of CFTs. One also can consider the family of so-called minimal models $\mathcal{M}(p,p^\prime)$, with $p,p^\prime$ integers, 
to build FQHE wavefunctions. While the Pfaffian state is related to $\mathcal{M}(4,3)$ which is the same as $Z_2$ Ising fermions\textcolor{red}{,}
the other states are different. The state at $\nu=2/5$ constructed from $\mathcal{M}(5,3)$ was first proposed by Simon et al.
in ref.(\onlinecite{Gaff1}) where it was called the ``Gaffnian''. It has certainly some desirable properties like 
good overlap with the true  
ground state for Coulomb interactions. Several studies have been devoted to its 
properties~\cite{Simon09,Flavin2011,Flavin2012,Seidel2010,Seidel2011,Chandran2011,SRR10,Wen2010}
However there is a potential problem which has been raised originally by Read~\cite{R1,R2,DRR}. One expects that the CFT from which one constructs
the FQHE trial wavefunction should be \textit{unitary} if the bulk state has a gap. This severely restricts minimal models
to $p=p^\prime+1$. In the plasma language~\cite{Bonderson2011} it means that there is no screening of charge.
Studies of Gaffnian quasiparticles have also raised doubts~\cite{TokeJain09}.
The Pfaffian and the $Z_k$ parafermions are all based on a unitary CFT hence qualify for
describing incompressible FQHE states. The other states should be gapless hence describe critical points presumably
in between several types of FQHE phases.

There is also another way of classifying the models based on the family of Jack polynomials~\cite{BH08,BHL08,BHL08bis,RBH09}.
These are multivariate symmetric polynomials that are known to play a role in several integrable systems.
It has been shown that they can be characterized by their vanishing properties when two or more
coordinates become equal. Some of the wavefunctions mentioned above, {\it i.e.}  the Laughlin, Moore-Read and parafermion wavefunctions\textcolor{red}{,} 
are all Jack polynomials. Generally speaking they are indexed by a real parameter and a partition of an  integer.
Bernevig and Haldane have proposed new trial FQHE states for filling factors $\nu =k/r$, $k$ and $r$ being integers.
They can be derived from a CFT which is the minimal model~\cite{BGS09,WJack} $\mathcal{W}_k(k+1,k+r)$. These models are unitary only if $r=2$
and this leads back to the Read-Rezayi family of states. We are thus facing again a set of wavefunctions that should describe critical states.
So it is natural to ask if standard tools used in the study of FQHE physics are able to directly prove the criticality of
these states. The original study of the Gaffnian state for example is rather inconclusive due to the small number of
particles reached in exact diagonalization studies as pointed out by the authors themselves.
Subsequent investigations have pointed out possible problems in the entanglement spectrum and also that there are level crossings
when interpolating between the Coulomb interaction and the special hard-core interaction for which the Gaffnian is the
exact ground state.

In this paper we concentrate on the Gaffnian state in order to obtain a direct evidence for criticality. We use
exact diagonalizations in the spherical geometry for the bosonic formulation and we are able to reach larger system sizes 
than previously considered.
The Hamiltonian we consider is a one-parameter family interpolating between the pure hard-core two-body delta function interaction
and the special three-body interaction involving derivatives of delta functions for which the Gaffnian is the exact ground state.
We start from the pure hard-core limiting case where there is ample evidence that the physics is gapped with a
description in terms of CF appropriate to bosonic systems. We observe then a definite scaling behavior of the gap 
of neutral collective excitations vs system size
which serves as a reference law when we tune the Hamiltonian towards the Gaffnian limit. Provided we focus on the right
angular momentum sector we observe a change in the scaling law which is fully consistent with a critical system.
The sphere geometry has no boundary and is well suited to get gap estimates. To get additional insight in the CFT properties
we also study the Gaffnian in the cylinder geometry with open boundary conditions. It is known that this is efficient
to capture edge state physics when adding a shallow confining potential. Of course we expect that in the thermodynamic limit
there should be no distinction between bulk and edge modes for a critical system. However it needs not be so
for finite size systems. Indeed we show that there are well-defined edge modes that can be classified according to the underlying
CFT. As in the case of the Pfaffian~\cite{Soule2013} the edge modes form conformal towers of the Virasoro algebra.
These special sets  of states are generated by primary operators of the CFT. In the Gaffnian case we find  a
new property~: the towers are ``heterotic'' i.e. the left and right moving modes are not generated by the same operator.
It is a very special combination of primaries that can explain the edge spectrum. While these properties are found
for all cylinder diameters, they become manifest in the thin-torus limit when the cylinder radius goes to zero. Then the problem
becomes purely electrostatic and can be exactly solved for a large number of particles.

In section II we discuss the vanishing properties of quantum Hall wavefunctions and their classifications. The Gaffnian state
is formulated explicitly. Section III is devoted to CFT aspects of the Gaffnian and we give elements of CFT concerning
conformal towers in the cylinder geometry and discuss edge mode properties. 
In section V we give gap estimates for a family of Hamiltonian and show evidence
for criticality of the Gaffnian state.
Finally section VI contains our conclusions.

\section{The Gaffnian wavefunction}
We first discuss FQHE wavefunctions in the planar disk geometry. In this case, the symmetric gauge is appropriate and
the lowest Landau level wavefunctions
are of the form~:
\be
\Psi(z_1,\dots,z_N)=P(z_1,\dots,z_N) {e}^{-\sum_i|z_i|^2/4},
\ee
where $P$ is a polynomial which is symmetric for bosons with coordinates $z_1,\dots,z_N$
and antisymmetric for fermions. In this paper we focus on spin-polarized states only. The magnetic length
is set to unity. The angular momentum $L_z$ with respect to the axis perpendicular to the plane is
a conserved quantity.
The Laughlin state for filling factor $\nu =1/m$ is given by~:
\be
P(z_1,\dots,z_N)=\prod_{i<j}(z_i-z_j)^m .
\label{Laughlin}
\ee
It is also very convenient from a theoretical point of view to put the particles on the surface of a sphere
with a radial magnetic field as if it were created by a magnetic monopole in the center. This geometry has
no boundaries and a finite area. The sphere has full rotation symmetry so states can be classified by their
total angular momentum $L$ in addition to the component $L_z$~: they form multiplets of definite $L$.
If the number of flux quanta piercing the sphere is $N_\phi$ then the LLL has degeneracy $N_\phi+1$
and a (unnormalized) basis can be taken as~:
\be
\Phi_S^M= u^{S+M} v^{S-M}, M=-S,\dots,+S,
\ee
where $2S=N_\phi$ and the spinors are given by $u=\cos(\theta/2) e^{i\phi/2}$, $v=\sin(\theta/2) e^{-i\phi/2}$
in spherical coordinates. Many-body wavefunctions then become polynomials in $u$,$v$ variables. For example
the Laughlin state is given by~:
\be
\Psi^{(m)}=\prod_{i<j}(u_i v_j-u_j v_i)^m .
\ee

The Laughlin wavefunction Eq.(\ref{Laughlin}) is a model state which is not an exact ground state for electrons with
Coulomb interactions. However it is known to be the exact ground state of a special hard-core Hamiltonian.
To understand this property in more detail, one has first to note that any two-body interaction Hamiltonian 
$\mathcal{H}_{2b}$
in the LLL
can be written as~:
\be
\hat{\mathcal{H}}_{2b}=\sum_{i<j} \sum_{k} V_k \hat{\mathcal{ P}}_{ij}^{(k)},
\ee
where $\hat{\mathcal{ P}}_{ij}^{(m)}$ is the projector onto relative angular momentum $m$ for the pair of particles $i,j$,
$m$ is a positive integer and the real numbers $V_m$ are the so-called Haldane pseudopotentials
that are defined by the choice of the Hamiltonian. In the Laughlin state Eq.(\ref{Laughlin}) each pair of particles
has a common factor $(z_i-z_j)^m$ and thus at least relative angular momentum $m$. We construct a special Hamiltonian
with the following recipe~: we take all pseudopotentials with $k>m$ equal to zero. Then the Laughlin state has exactly zero-energy
and if we ask for minimum angular momentum it is unique. For example fermions at $\nu =1/3$ have an exact Laughlin ground state
for the hard-core Hamiltonian $ \sum_{i<j}  V_1 \hat{\mathcal{ P}}_{ij}^{(1)}$.

There is a generalization~\cite{SRC07,SRC07bis} of this line of reasoning by considering now arbitrary $k$-body 
interactions instead of two-body interactions. We first discuss the three-body case $k=3$.
Similarly we can define relative angular momentum for three particles. For bosons the minimum value is then zero since all three
bosons may be in the same quantum state while for fermions it is 3. Consider now the projectors onto definite values of the three-body
relative angular momentum $\hat{\mathcal{ P}}_{ijk}^{(m)}$. They can be used to define hard-core Hamiltonians by projecting
out wavefunction components with relative momenta less than some value\textcolor{red}{s}. The simplest case corresponds to the projection
of the smallest value. This is the Hamiltonian~:
\be
\hat{\mathcal{H}}_{Pf}=\sum_{i<j<k} \hat{\mathcal{ P}}_{ijk}^{(m)}, m=0\, (\mathrm{bosons}), m=3\, (\mathrm{fermions}).
\label{PfH}
\ee
If we ask for the zero-energy ground state with minimal angular momentum there is a unique state which is called the
Moore-Read Pfaffian state. In planar coordinates the Bose Pfaffian can written as~:
\be
 \Psi_{Pf}^{Bose}=\mathcal{S}\,\,
\left[
\prod_{i_1<j_1}(z_{i_1}-z_{j_1})^2 
\prod_{i_2<j_2}(z_{i_2}-z_{j_2})^2 
\right].
\label{Pf1}
\ee
In this formula, we distribute the particles into two packets of equal size, the respective indices being $i_1,j_1,\dots$
in one packet and $i_2,j_2,\dots$ in the other one and we symmetrize ($\mathcal{S}$) over the choices of the two packets.
The vanishing properties of this wavefunction are easily read-off from this formula~: if two particles come close together
then the function does not vanish because they may belong to distinct packets, however if three particles coincide
then at least two of the them will be in the same packet and the wavefunction will vanish as the second power of their distance.
The filling factor is $\nu =1$ for the Bose case. It can also be written as~:
\be
\Psi_{Pf}^{Bose}={\rm Pf}\left(\frac{1}{z_i-z_j}\right)\prod_{i<j}\left(z_i-z_j\right) ,
\label{pfaffian}
\ee
where ${\rm Pf}$ stands for the Pfaffian symbol. This latter object is defined for an arbitrary
skew-symmetric $N\times N$ (N even) matrix $A$~:
\be
{\rm Pf}\left(A\right)=\sum_{\sigma} \epsilon_{\sigma} A_{\sigma(1)\sigma(2)}
A_{\sigma(3)\sigma(4)}...A_{\sigma(N-1)\sigma(N)},
\label{pfaffiandefinition}
\ee
where the sum
runs over all permutations of the index with N values and
$\epsilon_{\sigma}$ is the signature of the permutation. This is the original definition of the Pfaffian state
by Moore and Read from CFT arguments~\cite{Moore91}. The Pfaffian appears from the correlation functions of Ising model
Majorana fermions. The fermionic wavefunctions are obtained by multiplying by the Jastrow factor
$\prod_{i<j}(z_i-z_j)$ and the filling factor is now $\nu=1/2$.

Now we note that if we forbid relative angular momentum zero for three particles with the operator Eq.(\ref{PfH})
then the next allowed momentum is two. We can then consider the special Hamiltonian that projects out
these two possible three-body relative momenta~:
\be
\hat{\mathcal{H}}_{Gf}=\sum_{i<j<k} \hat{\mathcal{ P}}_{ijk}^{(0)}\, + \,
\sum_{i<j<k} \hat{\mathcal{ P}}_{ijk}^{(2)}\, .
\label{GfH}
\ee
This is the definition appropriate to the bosonic case. There is a unique zero-energy ground state
provided we ask for the smallest angular momentum. The wavefunction is then given by the following formula~:
\be
 \Psi_{Gf}=\mathcal{S}\,\,
\left[
\prod_{i_1<j_1}(z_{i_1}-z_{j_1})^{2+p} 
\prod_{i_2<j_2}(z_{i_2}-z_{j_2})^{2+p} 
\prod_{i_1<j_1}(z_{i_1}-z_{i_2})^{1+p} 
\prod\frac{1}{(z_{i_1}-z_{i_1+N/2})}
\right].
\label{Gf1}
\ee
It has been called the Gaffnian in Ref.\cite{Gaff1}. 
The Bose case corresponds to $p=0$ and the  filling factor is $\nu=2/3$ while $p=1$ describes a
fermionic state at $\nu=2/5$.
It is also possible~\cite{RGJ08} to write the Gaffnian wavefunction in the following way~:
\be
\Psi_{Gf}
=\prod_{i<j}(z_i-z_j)^{2(p+1)}\times
\mathcal{S}\,\,
\left[
\prod_{i_1<j_1}(z_{i_1}-z_{j_1})
\prod_{i_2<j_2}(z_{i_2}-z_{j_2})\,
{\rm per}\left[M\right]
\right].
\ee
in terms of the $N/2\times N/2$ matrix 
$M_{i,j} = \left[z_i-z_{j+N/2}\right]^{-1}$
and ${\rm per}\left[M\right]=\sum_{\{\sigma\}}\prod_{k=1}^N\,M_{k,\sigma(k)}$ is 
the permanent of  $M$, where the sum is over all $\sigma$ permutations of $N$ elements. 
 The symmetrized wave function (\ref{Gf1}), as a candidate for the FQHE at $\nu=2/5$ 
has been studied by Yoshioka {\sl et al.} in ref.(\onlinecite{Yoshioka88}), and one obtains a large overlap with the ED 
ground state for a Coulomb interaction. Furthermore, this wave function, also called 
"Gaffnian", has recently been studied within CFT and may support non-Abelian  
quasi-particle excitations.

\section{Conformal field theory construction of the Gaffnian}
We describe the construction of the Gaffnian wavefunction according to the general CFT approach
introduced by Moore and Read. The bulk wavefunction in the planar geometry is taken to be
the expectation value of a set of operators of a 1+1D CFT~:
\be
\Psi (z_1,\dots,z_N)=\langle 0|\mathcal{O}(z_1) \dots \mathcal{O}(z_N) \mathcal{O}_{bk}|0\rangle ,
\label{CFTDef}
\ee
where the operator $\mathcal{O}(z)$ is the product of a vertex operator for the charge sector
described by a free boson $\phi_c(z)$
and a statistical field belonging to some CFT $\mathcal{O}(z)=\psi(z) e^{i\phi_c(z)/\sqrt{\nu}}$.
One needs to add also a background charge operator $\mathcal{O}_{bk}$ to ensure global electric neutrality
and having a nonzero correlation function. Alternative constructions are possible with a different role for the neutrality~\cite{Milo1}.
In the case of the Laughlin state there is no statistical sector
and we just have to compute the correlation function of exponentials of a free boson~:
\be
\Psi_L (z_1,\dots,z_N)=\prod_{i<j}(z_i-z_j)^{1/\nu}.
\ee
In the Moore-Read Pfaffian case the statistical operator $\psi(z)$ is taken to be a Majorana fermion so that
there is an additional factor given by~:
\be
\langle \psi(z_1)\dots \psi(z_N)\rangle = {\rm Pf}\left(\frac{1}{z_i-z_j}\right).
\ee
This leads to the previous formula Eq.(\ref{pfaffian}) for the Pfaffian state.
Now if we consider a more general CFT state we need a field $\psi$ with fusion relation
$\psi\times\psi\sim \mathbf{1}$, conformal weight $\Delta_\psi$ and operator product expansion~:
\be
\psi(z)\psi(w)\sim \frac{1}{(z-w)^{2\Delta_\psi}}[\mathbf{1}+\dots],
\label{OPE}
\ee
so that the wavefunction is given by~:
\be
\Psi (z_1,\dots,z_N)=\langle \psi(z_1)\dots \psi(z_N)\rangle
\prod_{i<j}(z_i-z_j)^{2\Delta_\psi +q}.
\ee
The possible values of $q$ are then dictated by simple statistics requirements.
In the case of the Gaffnian state, authors of ref.(\cite{Gaff1}) have advocated the use
of the CFT defined by the Virasoro minimal model $\mathcal{M}(5,3)$ 
which contains a field $\psi$ with the correct fusion rule Eq.(\ref{OPE}) and dimension $\Delta_\psi=3/4$.
This CFT also contains two other primary fields $\varphi$ and $\sigma$ of dimensions
$\Delta_\varphi =1/5$ and $\Delta_\sigma =-1/20$ respectively.


According to the bulk/boundary correspondence we expect that the edge theory of a CFT-derived
quantum Hall state should be given by the very same CFT that is used to construct the bulk ground state
wavefunction. A precise derivation of this correspondence has been given in ref.(\onlinecite{DRR}).

If we consider the edge spectrum of a theory we expect the energies to be arranged in so-called conformal
towers of states since they should form a representation of the underlying Virasoro algebra of the CFT.
We briefly describe for completeness the tower structure which is explained in detail in the CFT literature~\cite{MalteBook,ginsparg}.
It is based on the use of the so-called generators of the Virasoro algebra $L_n$ with $n$ an integer, positive or negative.

If we create a state by acting with a primary operator onto the vacuum $|\Phi\rangle=\Phi(0)|0\rangle$ then this state is annihilated
by all Virasoro generators $L_k$ with $k>0$ and is an eigenstate of $L_0$ with eigenvalue $\Delta_\Phi$.
Action with the other generators with $k<0$ generates a family of states
called the descendants of the primary state~:
\be
|\{n_i\}\rangle =
L_{-n_k}\dots L_{-n_1}|\Phi\rangle ,
\label{Tower}
\ee
where the indices $n_i$ are positive. 
We call the ``level'' of such a state the integer $n\equiv\sum_k n_k$.
Not all these state are orthogonal and in a
given CFT there are relations between states at a given level so that the number of possible states
is not simply given by counting the partition of the level $n$ into $k$ integers.
If we introduce $p(\Phi ,n)$ as the number of independent states at level $n$, it is conveniently
manipulated through its generating functional~:
\be
\chi (\Phi , q) =\sum_{n=0}^{\infty}p(\Phi,n) q^n .
\label{char}
\ee
This quantity is called the character and encodes the structure of the representation of the Virasoro algebra.
Expansion in powers of $q$ allows to count the states. In the case of minimal models 
$\mathcal{M}(p,p^\prime)$ there is a finite number of primary operators $\Phi_{r,s}$
with $r=1,\dots p-1$, $s=1,\dots p^\prime-1$ with the redundancy $\Phi_{r,s}\equiv\Phi_{p-r,p^\prime-s}$.
Their scaling dimensions are given by~:
\be
h_{r,s}=\frac{(sp - rp^\prime)^2-(p-p^\prime)^2}{4pp^\prime}.
\ee
The Gaffnian CFT $\mathcal{M}(5,3)$ has fields $\psi = \Phi_{1,4}$, $\varphi=\Phi_{1,3}$,
$\sigma =\Phi_{1,2}$ and identity operator $\mathbf{1}=\Phi_{1,1}$.
To obtain the characters of a minimal model, we can use the
Rocha-Caridi formula which is valid for generic minimal models, unitary or not. It is
given by~:
\be
\chi(\Phi_{r,s}, q) = \frac{1}{(q)_\infty}
\sum_{k=-\infty}^{k=+\infty}(q^{k^2pp^\prime +k(pr-p^\prime s)}-q^{(kp^\prime +r)(kp+s)}),
\ee
where we have used the symbol~:
\be
{(q)_\infty}\equiv \prod_{n=1}^{\infty}(1-q^n).
\ee

Alternatively, a fermionic representation for the characters in the peculiar case of $\mathcal{M}(5,3)$
is available from the work of Kedem et al.~\cite{Kedem93}~:
\be
\chi(\Phi_{1,1}, q)=\sum_{f=0}^{\infty}\frac {q^{f(f+1)}}{(q)_{2f}},\quad
\chi(\Phi_{1,2}, q)=\sum_{f=0}^{\infty}\frac {q^{f^2}}{(q)_{2f}},
\ee
\be
\chi(\Phi_{1,3}, q)=\sum_{f=0}^{\infty}\frac {q^{f(f+1)}}{(q)_{2f+1}},\quad
\chi(\Phi_{1,4}, q)=\sum_{f=0}^{\infty}\frac {q^{f(f+2)}}{(q)_{2f+1}},
\ee
where~:
\be
{(q)_k}\equiv \prod_{n=1}^{k}(1-q^n).
\ee
With these character formula one can extract the degeneracy appearing in the towers. Low-lying state
counting is given in the upper part of Table I for each primary field of the Gaffnian CFT.
Provided the bulk is gapped, these degeneracies would be seen in the spectrum of edge modes
in the disk geometry. While this is the case for the Laughlin and Pfaffian case, we show in the next section 
that this is not case for the Gaffnian.

\begin{table}
\begin{tabular}{|l|l|l|}\hline
 $\Phi_{1,1}=\mathbf{1}$  & 1 0 1 1 2 2 4 4  6    7  10 11   \\ \hline
$\Phi_{1,2}=\sigma$           & 1 1 1 2 3 4 5 7  9   12 15 19             \\ \hline
$\Phi_{1,3}=\varphi$           & 1 1 2 2 3 4 6 7 10 12 16 20               \\ \hline
$\Phi_{1,4}=\psi$                & 1 1 1 2 2 3 4 5  7   9  11 14          \\ \hline\hline
$\phi_c +\mathbf{1} $                 & 1 1 3 5 10 16  29 45 74    \\ \hline
$\phi_c + \sigma$ & 1 2 4 8 15 26 44 72 115   \\ \hline
$\phi_c +\varphi $ & 1 2 5 9 17 29 50 80 129   \\ \hline
$\phi_c +\psi $ & 1 2 4 8 14 24 40 64 101   \\ \hline
\end{tabular} 
\label{chiraledge}
\caption{Upper part: number of states at each level of the Virasoro algebra for the statistical part $\mathcal{M}(5,3)$
of the Gaffnian CFT for each primary operator. This gives edge state counting in the disk geometry.
Lower part: we have added the boson sector to obtain the full Gaffnian state counting.}
\end{table}

\section{Gaffnian Conformal towers from the cylinder geometry}

In a generic non-chiral statistical critical model we expect energies to be related to Virasoro generators by~:
\be
\mathcal{H}\propto L_0+\bar{L}_0,
\ee
and momenta are expressed through the difference of the generators~:
\be
\mathcal{P}\propto L_0 - \bar{L}_0.
\ee
In these equations the barred Virasoro generators pertain to the antichiral copy of the algebra.
If we act with these relations onto a set of descendant states Eq.(\ref{Tower}), we generate a set
of energies and momenta~:
\be
E_{n,\bar{n}}\propto n+\bar{n}+h+\bar{h}, \quad P \propto n - \bar{n},
\label{ETower}
\ee
where $n=\sum_i n_i$ and $\bar{n}=\sum_i \bar{n}_i$ are levels in the chiral and anti-chiral Virasoro algebras.

In the realm of FQHE a non-chiral set-up is obtained by considering the cylinder geometry which has
naturally two counterpropagating edges. The Fock space of the edge theory is given by the tensor
product of the two boundary Fock spaces. We use thus use periodic boundary conditions along one
direction of the cylinder and impose for simplicity a hard cut-off in orbital space to define a finite-dimensional problem.
In this geometry there is only one conserved momentum $\mathcal{K}$ along the periodic direction. 
It can be used to label many-body eigenstates.

The Gaffnian Hamiltonian in the disk geometry can be written as~:
\be
\hat{\mathcal{H}}_{Gf}=\sum_{i<j<k} \mathcal{S}\left[\,\nabla^4_i\,
\delta^2(z_i-z_j)\, \delta^2(z_j-z_k)\right].
\label{GHd}
\ee

\begin{figure}[htb]
\includegraphics[width=0.5\columnwidth]{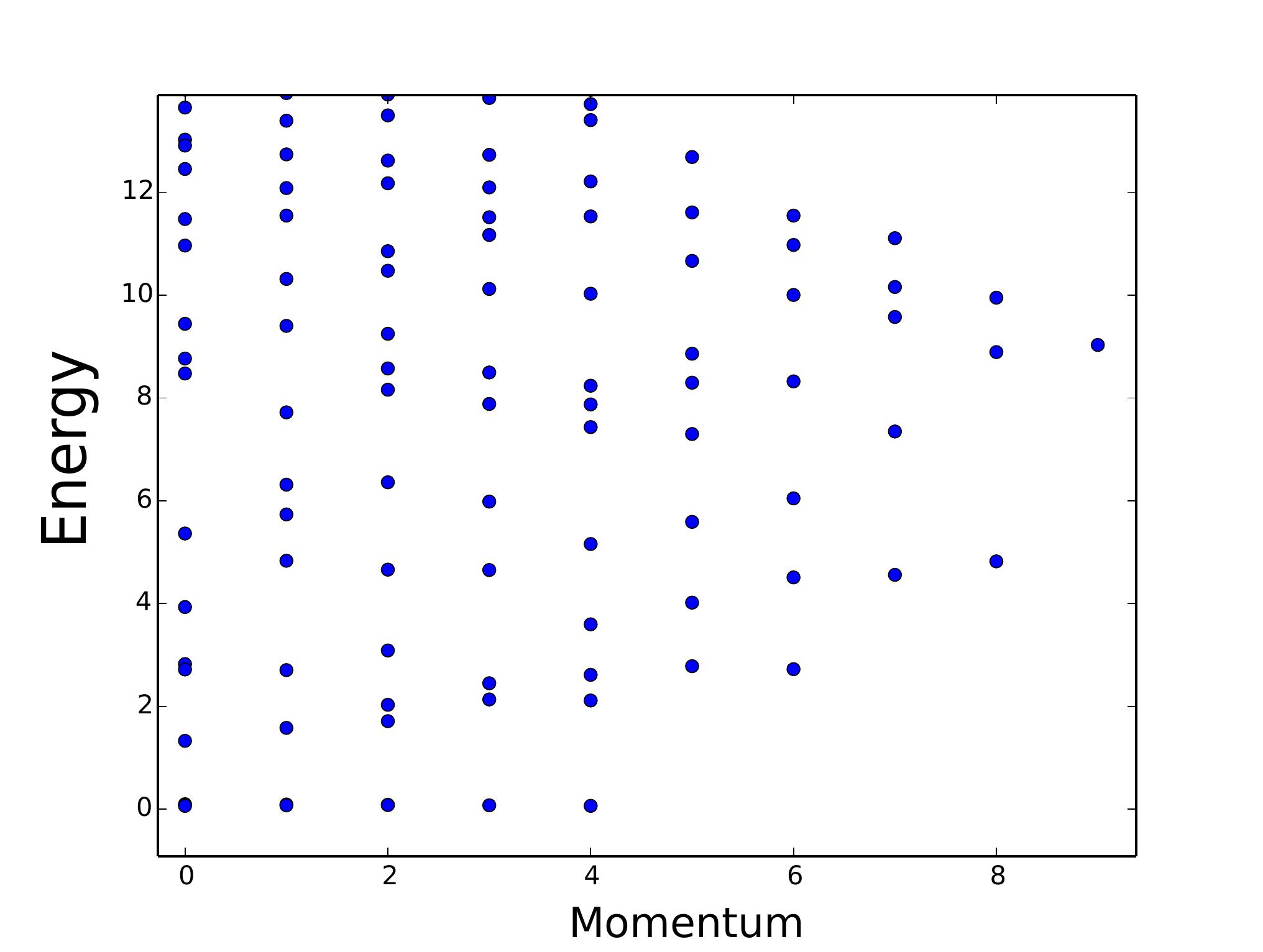}
\caption{Low-lying spectrum of the special Gaffnian Hamiltonian Eq.(\ref{GHd}) in the cylinder geometry
for N=8 bosons and 10 orbitals. Eigenstates are classified as a function of the conserved total momentum $K$.
There is a set of zero-energy eigenstates that extend up to $K=4$  that are Gaffnian quasiholes.}
\label{Gaff8}
\end{figure}

This Hamiltonian when written in the cylinder basis has a unique ground state with zero momentum
provided we fix the number of orbitals as~:
\be
2\mathcal{K}+1 = \frac{3}{2} N - 2,
\ee
as in the spherical geometry. If we add extra orbitals then additional zero-energy quasiholes appear.
This is displayed in Fig.(\ref{Gaff8}). 

%

\begin{figure}[htb]
\includegraphics[width=0.8\columnwidth]{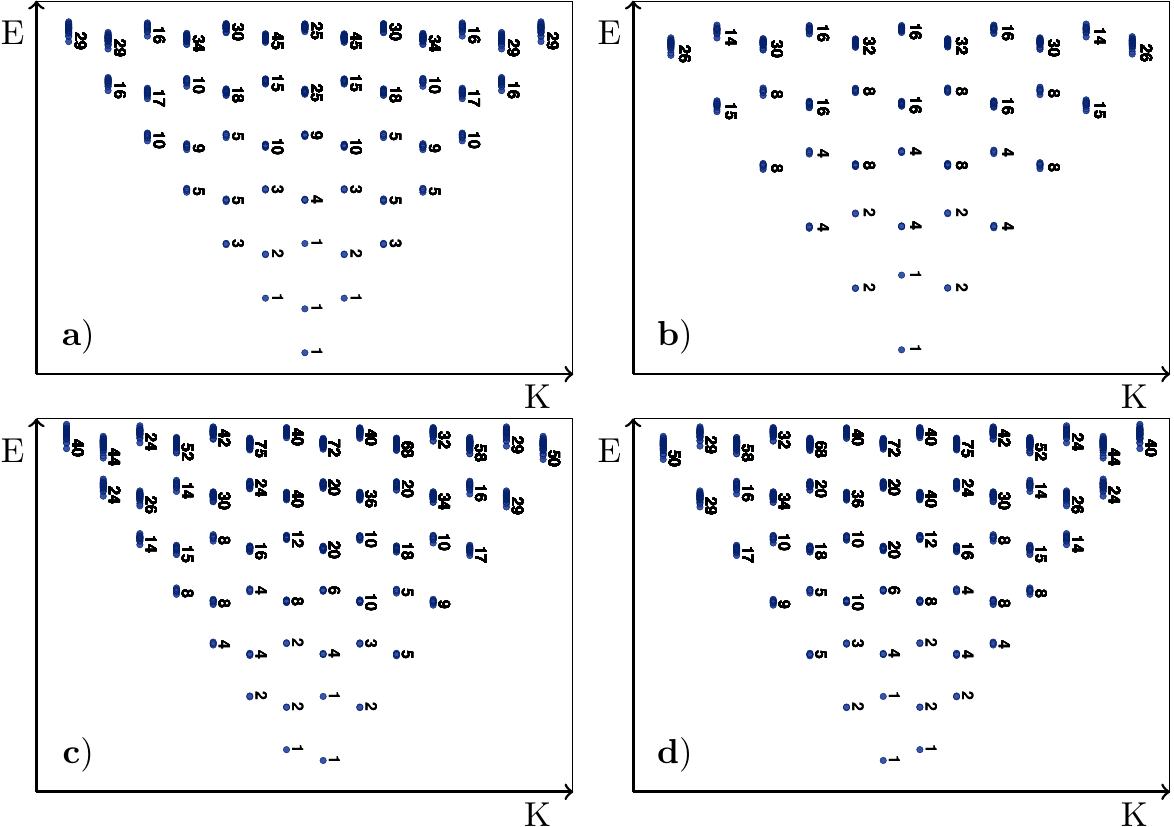}
\caption{Conformal towers of the Gaffnian state in the thin torus limit. The spectra are found for even number of bosons 
in panels (a) and (b), odd number of bosons for (c) and (d). Operator assignments of these towers are given in the text.
We note that odd towers (c) and (d) involve different operators for left and right moving modes~:
they are ``heterotic''.}
\label{Towers}
\end{figure}

The set of quasihole states is exactly degenerate provided there is perfect
translation invariance. If we now add a shallow confining potential along the axis of the cylinder the degeneracy
is lifted and it is possible to identify the edge excitations as conformal towers in the low-energy points of the spectrum.
This is possible in the Laughlin~\cite{Soule2012} and Pfaffian~\cite{Soule2013} cases. 
This picture is valid as long as there is
a clear separation between the set of quasihole-derived states and the bulk states. In the Gaffnian case we show in the next  section
that the bulk gap goes to zero in the thermodynamic limit. This means that the edge mode identification should be considered
as an artifact~: they will mix with bulk modes as soon as the system is large enough. Indeed we observe that when the radius
$L$ of the cylinder is of order of a few magnetic lengths it is difficult to identify the edge modes. The situation becomes on the contrary
very clear when going to the thin torus limit $L\rightarrow 0$. It is then possible to study numerically large systems because
one has just to solve a problem of minimum electrostatic energy under constraints.
The lowest-lying state are then clearly arranged into conformal towers
as can be seen in Fig.(\ref{Towers}). The ground state in each of the sectors we find can be tentatively identified
by comparing the degeneracies observed with those deduced from the CFT predictions in the lower part of table (II)
which includes the charged boson mode in addition to the statistical field.

When the number of bosons is even there are two towers that alternate when increasing the total momentum~:
(a) and (b) in Fig.(\ref{Towers}). The tower of type (a) is generated by the identity operator as well as $\varphi$ operator.
The ground state of tower (a) is thus~:
\be
|0,a\rangle = \mathbf{1}\,\, e^{2in\phi_c/\sqrt{6}}e^{2in\bar{\phi}_c/\sqrt{6}}|0\rangle,
\ee
and the first excited state with the same total momentum is obtained by acting with the primary field $\varphi$~:
\be
|1,a\rangle = 
 \varphi \bar{\varphi}\,\,
e^{2in\phi_c/\sqrt{6}}e^{2in\bar{\phi}_c/\sqrt{6}}|0\rangle .
\ee
For tower (b) it involves the two other primary fields, the ground state is now~:
\be
|0,b\rangle = \sigma\bar{\sigma}\,\, e^{i(2n+1)\phi_c/\sqrt{6}}e^{i(2n+1)\bar{\phi}_c/\sqrt{6}}|0\rangle,
\ee
and the first excited state now given by~:
\be
|1,b\rangle = \psi \bar{\psi}\,\,
e^{i(2n+1)\phi_c/\sqrt{6}}e^{i(2n+1)\bar{\phi}_c/\sqrt{6}}|0\rangle .
\ee
These towers are analogous to those found in the Pfaffian case~\cite{Soule2013} in the sense
that the left and right moving modes are generated by the same operator. In these formulas $n$ is an integer
that varies as we go from one tower to its neighbor. Increasing total momenta and hence going through several
towers means that we are encountering all the topological sectors~\cite{Ardonne09,BGS09} of the Gaffnian (in general multiple times).

When there is an odd number of bosons there are also two alternating types of towers~: see (c) and (d)
in Fig.(\ref{Towers}) that are mirror symmetric from each other. Hence there is essentially only one type of conformal tower.
While their almost twofold ground state degeneracy is reminiscent of the $\psi$ tower of the Pfaffian,
they are structurally different since now right and left moving modes are not generated by the same operator.
To borrow terminology from string theory~\cite{GSW} they are aptly said to be ``heterotic''.
In the (c) case the leftmost ground state leads to a tower based on two operators 
$\psi$  for the left-handed modes and the identity for the right-handed modes..
Its almost degenerate rightmost ground state generates a tower based 
on $\sigma$ for the left modes and $\varphi$ for the right modes.
We have checked that these assignments hold at least up to the sixth level.
This can be easily checked with the data in Table I.
The states that appear in these towers are bona fide quantum mechanical states with
a perfectly well-defined positive scalar product. Since the underlying CFT is non-unitary it means
that this structure cannot persist in the true thermodynamic limit. On the cylinder geometry,
we observe that when going to large cylinder radius it is no longer possible to identify clearly the descendant states.
This is a hint of what goes wrong with the Gaffnian. In fact we now give direct evidence in the next section
for its critical character.

\section{Gap estimates on the spherical geometry}

\begin{figure}[htb]
\includegraphics[width=0.4\columnwidth]{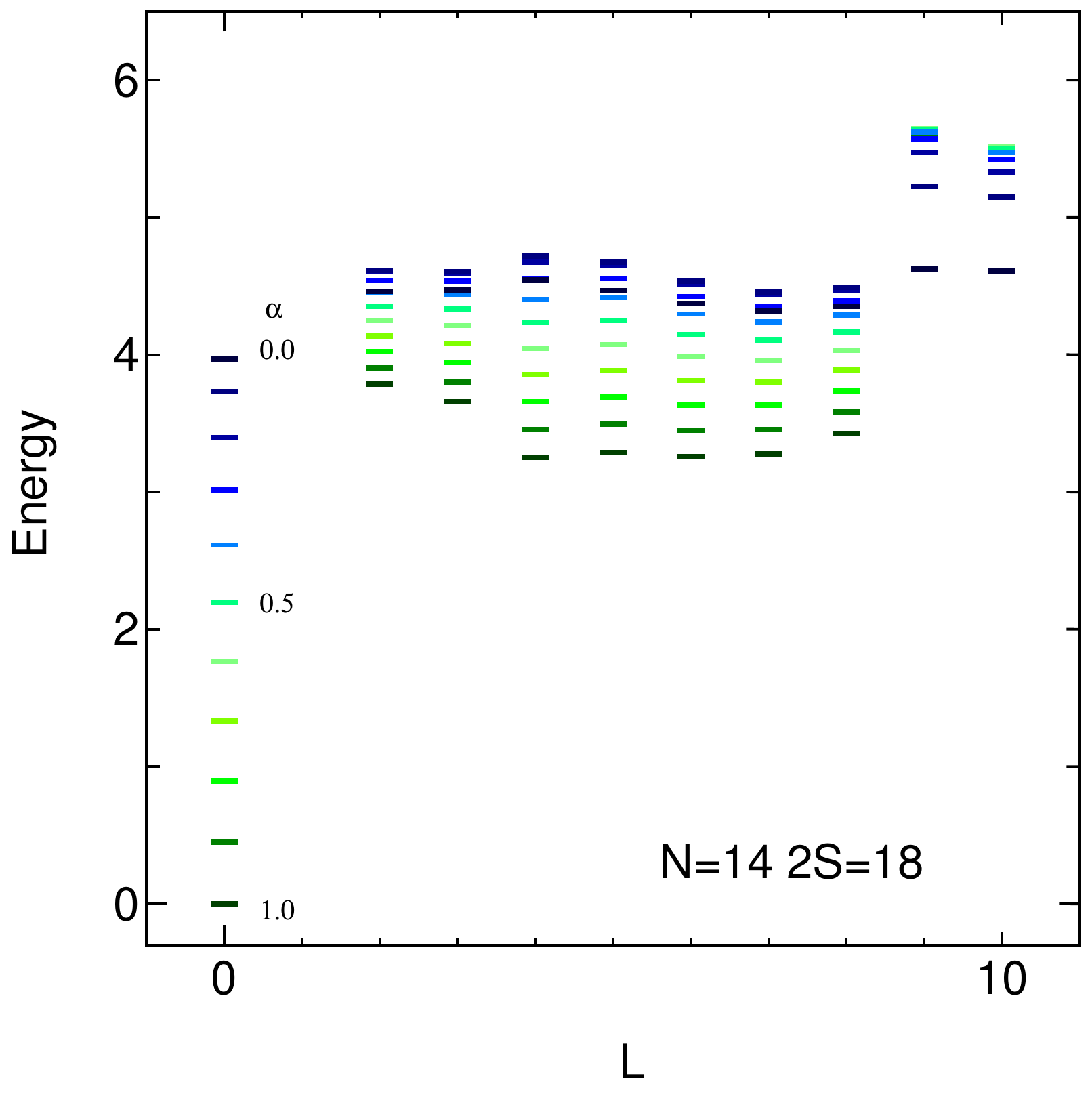}
\caption{The energy spectrum of the family of Hamiltonians interpolating between the Gaffnian and the pure
two-body projector. There are N=14 bosons on the sphere with $N_\phi=18$.
Energy levels are classified by their total angular momentum in the spherical geometry.
The two-body case is at the top of the picture and displays a collective mode with a dispersion relation similar to 
what is observed in smaller systems. When we switch on the three-body Hamiltonian Eq.(\ref{GfH}) the dispersion relation
first flattens and then acquires a hanging-chain shape as in other paired states. The value of $L_{tot}$ with minimal
energy changes as a function of $\alpha$.}
\label{Spectrum}
\end{figure}
To investigate the possible criticality of the Gaffnian state we revert to the spherical geometry.
We focus on its bosonic formulation because this allows for larger sizes to be studied numerically.
When written on the sphere the wavefunction Eq.(\ref{Gf1}) requires a definite relationship between flux
and number of bosons~:
\be
2S=N_\phi=\frac{3}{2}N - 3.
\label{shift}
\ee
This coincides with the relation for the principal  or Jain series of quantum Hall fractions at filling factor
$\nu=2/3$. Indeed it is known that bosons in the LLL with hard-core interactions form usual Abelian
FQHE states at fillings $\nu=p/(p+1)$. They fit in the CF scheme by introducing $^1$CF entities
that are bound states of one boson and one vortex. These $^1$CFs feel reduced magnetic field
$2S^*=2S-(N-1)$ and integer fillings of CF LLs describes a FQHE state as in the fermionic FQHE.
With two filled levels we obtain the Bose state $\nu=2/3$ which appears to be incompressible in numerical studies
for the pure contact interaction. On the sphere such a state appears as a rotationally invariant $L_{tot}=0$
state. Neutral excitations are obtained by promoting some CF from full to empty levels. This operation creates
 a branch of states dubbed excitons where a CF goes from the level with $L=S^*+1$ to
the first empty level $L=S^*+2$. This branch is prominent in exact diagonalization studies and is reasonably well
described by CF wavefunctions adapted for the Bose statistics.

\begin{figure}[htb]
\includegraphics[width=0.4\columnwidth]{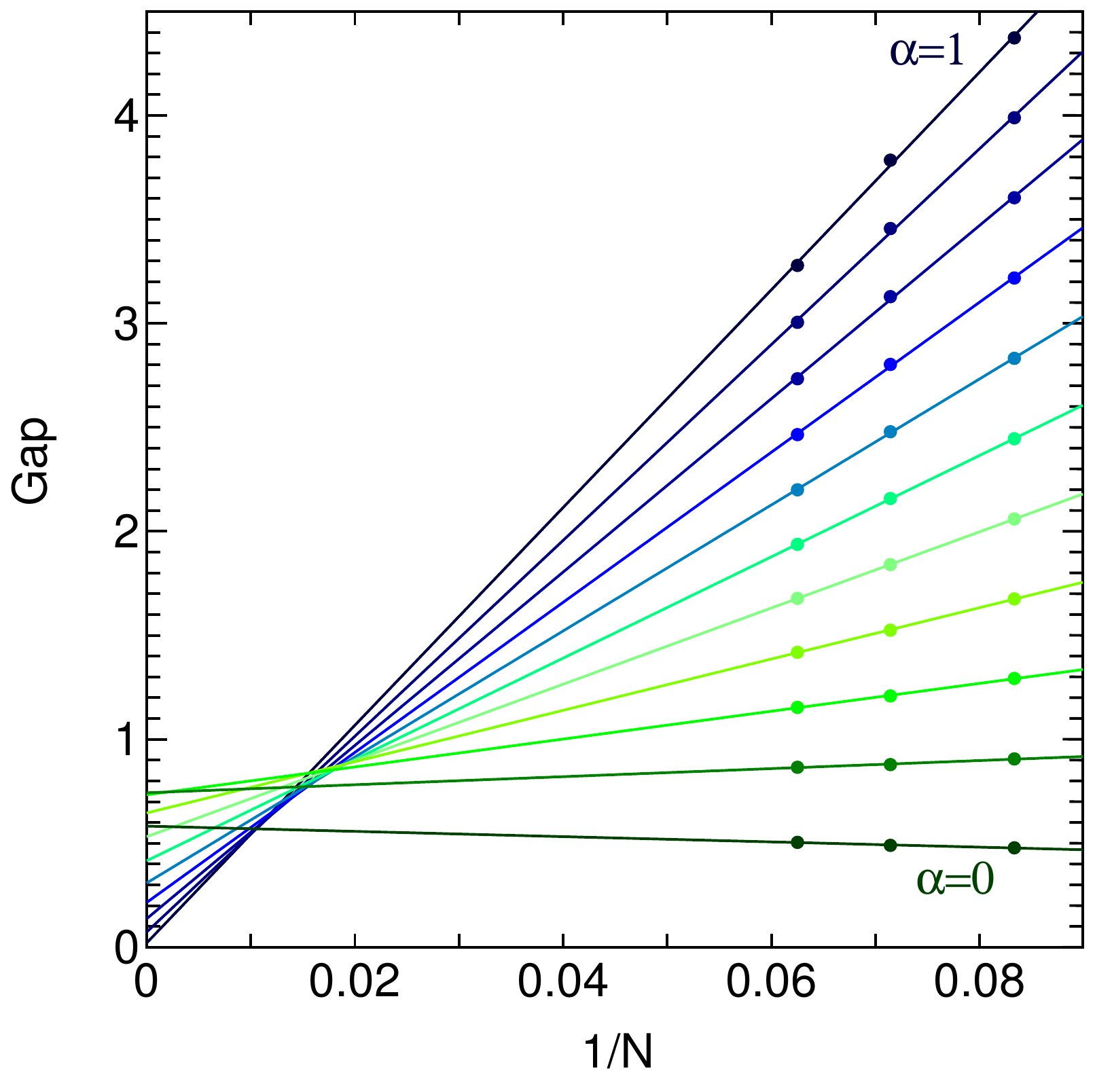}
\caption{Extrapolation of gaps to the thermodynamic limit. The gaps are plotted versus inverse of the
number of particles. We display only the largest sizes N=12,14,16 bosons. The angular momentum of the state
is fixed at $L_{tot}=2$. Only the special Gaffnian Hamiltonian is gapless within the precision of our method.
The lowest-lying straight line corresponds to $\alpha=0$ and the topmost line to $\alpha=1$
with intermediate values differing by 0.1.}
\label{Scaling}
\end{figure}

In order to have some control on the gap estimate for the Gaffnian we study a family of Hamiltonians
that interpolates between the special three-body interaction of Eq.(\ref{GfH}) and the pure two-body hard-core
case~:
\be
\hat{\mathcal{H}}_{\alpha}=
\alpha\,\, \hat{\mathcal{H}}_{Gf}+(1-\alpha)\,\sum_{i<j} \hat{\mathcal{ P}}_{ij}^{(0)},
\label{interpol}
\ee
At fixed flux-number of particles relation Eq.(\ref{shift}) one interpolates between the Gaffnian state and the CF state.
The spectra for N=14 bosons is displayed in Fig.(\ref{Spectrum}) for various values of the $\alpha$ interpolating
parameter. 
We have plotted only the  lowest energy state in each $L$ sector. 
The ground state is always at $L=0$.
It is only for $\alpha =1$ that the ground state energy is zero. For other values there are pairs of particles with zero
relative angular momentum. If we now try to extrapolate the finite-size gap to the thermodynamic limit, we observe
the scaling in Fig.(\ref{Scaling}). Taking into account only the three largest sizes, we find very good regularity.
The pure hard-core two-body interaction has only small size dependence and leads to a rather precise gap estimate.
For increasing values of $\alpha$ towards the pure Gaffnian Hamiltonian we find also a rather smooth dependence
that we consider as evidence for zero gap when $\alpha=1$.

\begin{figure}[htb]
\includegraphics[width=0.4\columnwidth]{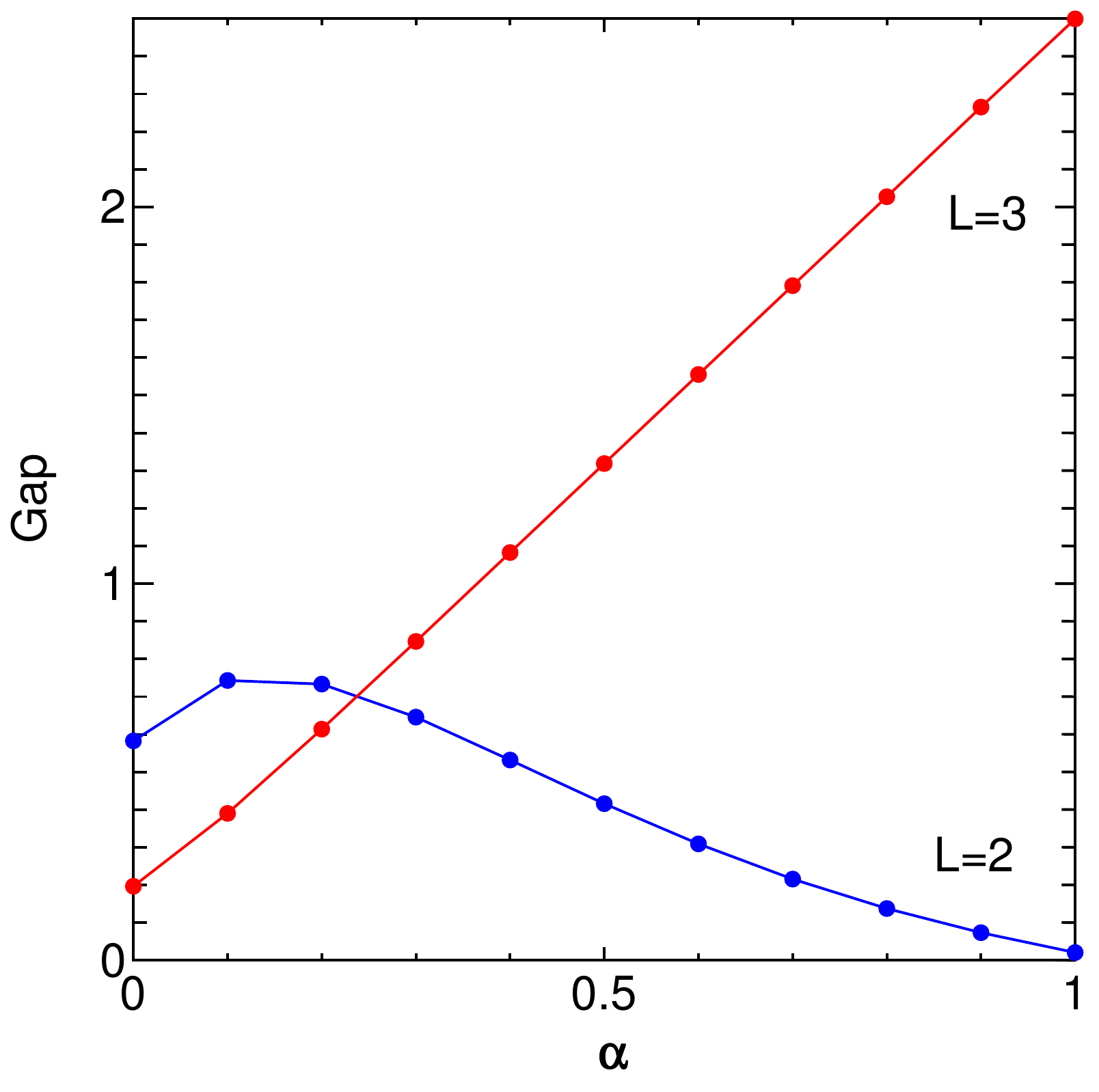}
\caption{The extrapolated gaps as a function of the interpolating parameter $\alpha$
based on sizes $N=12,14,16$.
The lower (blue online) curve is obtained from $L_{tot}=2$ gaps and the upper (red online)
curve is for $L_{tot}=3$. The gapped two-body case is for $\alpha =0$ and the Gaffnian case for
$\alpha=1$. There is evidence for zero gap only at the Gaffnian point and the gap opens linearly
with $|1-\alpha|$.}
\label{GapEvolution}
\end{figure}

We present estimates for infinite-system gaps in Fig.(\ref{GapEvolution}). The gap opens immediately as soon as we add
the two-body interaction. The gap behaviour is close to linear as a function of $|1-\alpha|$ close to $\alpha=1$. This means that
the two-body interaction should be considered as a relevant perturbation with respect to the critical Gaffnian state.
The gap in the  $L=3$ sector is also plotted in Fig.(\ref{GapEvolution}) and shows that there are in fact multiple
level crossings in the excited states when approaching the CF regime for $\alpha\approx 0$.

This is similar to the observation of ref.(\onlinecite{TokeJain09}) of level crossings in the sector of charged quasiparticles.

\section{Conclusions}

We have studied various aspects of the Gaffnian state which is based on a non-unitary CFT $\mathcal{M}(5,3)$.
As such it is expected to describe a gapless state of matter~\cite{DRR}.  
By exact diagonalizations on large systems in the spherical geometry
we have given numerical evidence for its critical nature. The pure two-body contact interaction projected onto
the LLL appears to be a relevant perturbation which opens a gap even for infinitesimal coupling.
The criticality of the Gaffnian we find is in agreement with other lines of attack~\cite{RBH09,Wu,BBR12}.

By use of the cylinder geometry we have shown the appearance of the conformal towers expected from
its underlying CFT structure. The towers are based on all primary operators of the theory and in the
case of an odd number of particles they are ``heterotic'' i.e. their left and right movers are created
by different operators, at variance with the Laughlin~\cite{Soule2012} or Moore-Read case~\cite{Soule2013}.
These towers appear however only in the thin-torus limit which has no thermodynamic limit.
Indeed since there are negative norm states in the representations of the Virasoro algebra, there should be drastic changes
in the energy spectrum when going to the true thermodynamic limit. Here what happens is that there are states that were
looking like ``bulk'' states in Fig.(\ref{Spectrum}) that mix with ``edge'' states.

\begin{acknowledgments}
We thank P. Soul\'e for collaboration at an early stage of this work.
We wish to acknowledge discussions  with J. B. Zuber as well as a careful reading of the manuscript. 
Part of the numerical calculations were performed thanks an IDRIS-CNRS compuer time
allocation under project number 100336.

\end{acknowledgments}



\begin{thebibliography}{99}

\bibitem{Laughlin83}
R. B. Laughlin,
Phys. Rev. Lett. {\bf 50}, 1395 (1983).

\bibitem{JainBook}
J. K. Jain,
``Composite Fermions'', Cambridge University press, New York (2007).

\bibitem{Moore91}
G. Moore and N. Read,
Nucl. Phys. B{\bf 360}, 362 (1991).

\bibitem{RMP08}
C. Nayak, S. H. Simon, A. Stern, M. Freedman, and S. Das Sarma,
Rev. Mod. Phys. {\bf 80}, 1083 (2008).

\bibitem{Kumar2010}
A. Kumar, G. A. Csathy, M. J. Manfra, L. N. Pfeiffer, and K. W. West,
Phys. Rev. Lett. {\bf 105}, 246808 (2010).

\bibitem{RR}
N. Read and E. H. Rezayi,
Phys. Rev. B{\bf 59}, 8084 (1999).

\bibitem{Regnault03}
N. Regnault and Th. Jolicoeur,
Phys. Rev. Lett. {\bf 91}, 030402 (2003);
Phys. Rev. B{\bf 69}, 235309 (2004).

\bibitem{Chang05}
C.-C. Chang, N. Regnault, Th. Jolicoeur, and J. K. Jain,
 Phys. Rev. A{\bf 72}, 013611 (2005).

\bibitem{Gaff1}
S. H. Simon, E. H. Rezayi, N. R. Cooper, and I. Berdnikov,
Phys. Rev. B{\bf 75}, 075317 (2007).

\bibitem{ginsparg}
 P. Ginsparg, ``Applied Conformal Field Theory'', Les Houches lectures session XLIX,
published in ``Fields, Strings, and Critical Phenomena'', ed. by E. Brezin and J. Zinn-Justin,
North-Holland, Elsevier (Amsterdam, 1989).

\bibitem{MalteBook}
Malte Henkel, 
``Conformal Invariance and Critical Phenomena'', 
Springer Verlag, Berlin (1999).

\bibitem{Simon09}
S. H. Simon,
Journal of Physics A: Mathematical and Theoretical {\bf 42}, 055402 (2009).

\bibitem{Flavin2011}
J. Flavin and A. Seidel,
Phys. Rev. X{\bf 1}, 021015 (2011).

\bibitem{Flavin2012}
J. Flavin, R. Thomale, and A. Seidel,
Phys. Rev. B{\bf 86}, 125316 (2012).

\bibitem{Seidel2010}
A. Seidel,
Phys. Rev. Lett. {\bf 105}, 026802 (2010).

\bibitem{Seidel2011}
A. Seidel and K. Yang,
Phys. Rev. B{\bf 84}, 085122 (2011).

\bibitem{Chandran2011}
A. Chandran, M. Hermanns, N. Regnault, and B. A. Bernevig,
Phys. Rev. B{\bf 84}, 205136 (2011).

\bibitem{SRR10}
S. H. Simon, E. H. Rezayi, and N. Regnault,
Phys. Rev. B{\bf 81}, 121301 (2010).

\bibitem{Wen2010}
Y.-M. Lu, X.-G. Wen, Z. Wang, and Z. Wang,
Phys. Rev. B{\bf 81}, 115124 (2010).

\bibitem{R1}
N. Read,
Phys. Rev. B{\bf 79}, 245304 (2009).

\bibitem{R2}
N. Read,
Phys. Rev. B{\bf 79}, 045308 (2009).

\bibitem{DRR}
J. Dubail, N. Read, and E. H. Rezayi,
Phys. Rev. B{\bf 86}, 245310 (2012).

\bibitem{Bonderson2011}
P. Bonderson, V. Gurarie, and C. Nayak,
Phys. Rev. B{\bf 83}, 075303 (2011).

\bibitem{TokeJain09}
C. T\"{o}ke and J. K. Jain,
Phys. Rev. B{\bf 80}, 205301 (2009).

\bibitem{BH08}
B. A. Bernevig and F. D. M. Haldane,
Phys. Rev. B{\bf 77}, 184502 (2008).

\bibitem{BHL08}
B. A. Bernevig and F. D. M. Haldane,
Phys. Rev. Lett. {\bf 101}, 246806 (2008).

\bibitem{BHL08bis}
B. A. Bernevig and F. D. M. Haldane,
Phys. Rev. Lett. {\bf 100}, 246802 (2008).

\bibitem{RBH09}
N. Regnault, B. A. Bernevig, and F. D. M. Haldane,
Phys. Rev. Lett. {\bf 103}, 016801 (2009).

\bibitem{BGS09}
B. A. Bernevig, V. Gurarie, and S. H. Simon,
Journal of Physics A: Mathematical and Theoretical {\bf 42}, 245206 (2009).

\bibitem{WJack}
B. Estienne and R. Santachiara,
J. Phys. A: Math. Theor. {\bf 42},  445209 (2009).

\bibitem{Soule2013}
P. Soul\'e, Th. Jolicoeur, and Ph. Lecheminant,
Phys. Rev. B{\bf 88}, 235107  (2013).

\bibitem{SRC07}
S. H. Simon, E. H. Rezayi, and N. R. Cooper,
Phys. Rev. B{\bf 75}, 075318 (2007).

\bibitem{SRC07bis}
S. H. Simon, E. H. Rezayi, and N. R. Cooper,
Phys. Rev. B{\bf 75}, 195306 (2007).

\bibitem{RGJ08}
N. Regnault, M. O. Goerbig, and Th. Jolicoeur,
Phys. Rev. Lett. {\bf 101}, 066803 (2008).

\bibitem{Yoshioka88}
D. Yoshioka, A. H. MacDonald, and S. M. Girvin, 
Phys. Rev. B {\bf 38}, 3636 (1988).

\bibitem{Milo1}
M. V. Milovanovic, Th. Jolicoeur, and I. Vidanovic,
Phys. Rev. B{\bf 80}, 155324 (2009).

\bibitem{Kedem93}
R. Kedem,  T. R. Klassen, B. M. McCoy, and E. Melzer,
Phys. Lett. B{\bf 307}, 68 (1993).

\bibitem{Soule2012}
P. Soul\'e and Th. Jolicoeur,
Phys. Rev. B{\bf 86}, 115214  (2012).

\bibitem{Ardonne09}
Eddy Ardonne,
Phys. Rev. Lett. {\bf 102}, 180401 (2009).

\bibitem{GSW}
M. B. Green, J. H. Schwarz, and E. Witten,
``Superstring theory'', p. 305,
Cambridge University Press, Cambridge (1987).

\bibitem{Wu}
Yang-Le Wu, B. Estienne, N. Regnault, and B. Andrei Bernevig,
``Braiding non-Abelian quasiholes in fractional quantum Hall states'',
e-print arXiv:1405.1720

\bibitem{BBR12}
B. A. Bernevig, P. Bonderson, and N. Regnault,
``Screening Behavior and Scaling Exponents from Quantum Hall Wavefunctions'',
e-print arXiv:1207.3305

\end{thebibliography}
\end{document}